\documentclass[preprintnumbers,aps,amsmath,amssymb,pra,twocolumn,showpacs,superscriptaddress]{revtex4-1}
\usepackage[dvipdfm]{graphicx}

\newcommand{\ket}[1]{|#1\rangle}
\newcommand{\bra}[1]{\langle#1|}

\newcommand{\Epsilon}{\mathcal{E}}

\newtheorem{lemma}{\textit{Lemma}}

\begin{document}



\title{Quantum computation over the butterfly network}



\author{Akihito Soeda}
\affiliation{Department of Physics, Graduate School of Science,
University of Tokyo, Tokyo 113-0033, Japan}

\author{Yoshiyuki Kinjo}
\affiliation{Department of Physics, Graduate School of Science,
University of Tokyo, Tokyo 113-0033, Japan}

\author{Peter S. Turner}
\affiliation{Department of Physics, Graduate School of Science,
University of Tokyo, Tokyo 113-0033, Japan}

\author{Mio Murao}
\affiliation{Department of Physics, Graduate School of Science,
University of Tokyo, Tokyo 113-0033, Japan}
\affiliation{Institute for Nano Quantum Information Electronics,
University of Tokyo, Tokyo 153-8505, Japan}



\date{\today}

\begin{abstract}
In order to investigate distributed quantum computation under restricted network resources, we introduce a quantum computation task over the butterfly network where both quantum and classical communications are limited.   We consider deterministically performing a two-qubit global unitary operation on two unknown inputs given at different nodes, with outputs at two distinct nodes.   By using a particular resource setting introduced by M. Hayashi [Phys. Rev. A \textbf{76}, 040301(R) (2007)], which is capable of performing a swap operation by adding two maximally entangled qubits (ebits) between the two input nodes, we show that unitary operations can be performed without adding any entanglement resource, if and only if the unitary operations are locally unitary equivalent to controlled unitary operations.  Our protocol is optimal in the sense that the unitary operations cannot be implemented if we relax the specifications of any of the channels. We also construct protocols for performing controlled traceless unitary operations with a 1-ebit resource and for performing global Clifford operations with a 2-ebit resource.
\end{abstract}


\maketitle


\section{Introduction}

Distributed quantum computation aims to perform a large-scale quantum computation using a collection of smaller scale quantum computers connected by communication channels. There are several distributed quantum computation architectures proposed for different purposes \cite{DQC}.  In general, distributed computation can be modeled by a combination of computation at each node and communication between the nodes, for both the quantum and classical cases.  For distributed quantum computation, initially shared entanglement among the nodes can be used as a resource, as well as quantum and classical communication channels.  The amount of communication between the nodes required to perform quantum computation tasks has been analyzed by quantum communication complexity theory \cite{QComC}.

\begin{figure}
\includegraphics[width=5cm]{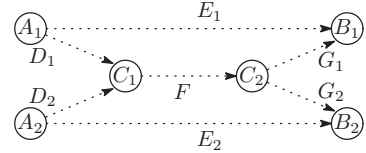}%
\caption{The butterfly network (drawn horizontally). The 2-pair communication
 problem aims to transmit information (bit or qubits) from $A_1$ to $B_2$ and
 from $A_2$  to $B_1$ concurrently via nodes $C_1$ and $C_2$.
  The directed edges $D_1$, $D_2$, $E_1$, $E_2$, $F$, $G_1$ and $G_2$ denote
 communication channels.  The channel $F$ exhibits the bottleneck.}
\label{butterfly}
\end{figure}

As the ``distributedness'' of a quantum computation increases, the scale
 (i.e., the number of qubits) of the quantum computer at each node decreases, while the number of nodes increases.
   The communication resources (quantum channels, classical channels, and shared
entanglement) form an increasingly large network and the amount of
 communication required grows.
   In any such large network, one will inevitably be faced with a bottleneck
 problem, where communication capacities in some region are lower than that
 required by a straightforward implementation of the protocol.
   This bottleneck restricts the total performance of communication.
   In network information theory, this problem has been extensively studied
 for the last decade or so under the name network source coding \cite{Qi:200}.
 Although solving general network problems is difficult, a solution of the
 2-pair communication (communications of two disjoint sender-receiver pairs)
 bottleneck problem is known for a simple directed network called the
 {\it butterfly network} \cite{Ajlswede}  (shown in Fig.~\ref{butterfly}) in the classical case.

In the quantum case, where the no-cloning theorem holds, the method used in the
 classical case cannot be applied directly, since it involves cloning inputs.
   Nevertheless, in \cite{Hayashi et al}, it is shown that efficient network
 source coding on the quantum butterfly network, where edges represent 1-qubit
 quantum channels,  is possible for transmitting approximated states.
   Asymptotic rates of high fidelity quantum communication have been obtained
 for various networks including the butterfly network with and without
 additional entanglement \cite{Debbie}.
   In \cite{Hayashi}, it is shown that perfect quantum 2-pair communication
 over the butterfly network is possible if we add two maximally entangled
 qubits (ebits) between the inputs and allow each channel (edge) to use either
 1 qubit of communication or 2 (classical) bits of communication.  Recently it
 has been shown that if we allow free classical communication between all
 nodes, perfect 2-pair communication over the butterfly network is possible
 without additional resources \cite{Kobayashi}.

In this paper, we investigate the performance of efficient distributed quantum
 computation over such bottlenecked networks where both quantum and classical
 communication is restricted.
   We combine both quantum computation, namely, performing a gate operation on
 inputs, and network communication, namely, sending outputs, in a
{\it single task}.
   The task we consider is to deterministically implement a global unitary
 operation on two inputs at distant nodes and obtain two outputs at distinct
 nodes connected by the particular butterfly network introduced by Hayashi
 \cite{Hayashi}.
We show that unitary operations can be performed without adding any entanglement
 resource, if and only if the unitary operations are locally unitary equivalent to controlled unitary operations, by constructing a protocol for sufficiency and analyzing entangling capability of the butterfly network for necessity.  Further, we prove that our protocol is optimal in terms of resource usage.
 We also present constructions of protocols for performing controlled
 traceless unitary operations with a 1-ebit resource and for performing global
 Clifford operations with a 2-ebit resource.

The rest of the paper is organized as follows.
   In Sec. \ref{SecHay},  we introduce our task of implementing a global
 unitary operation over a network, and review Hayashi's protocol \cite{Hayashi}
 in the context of implementing a swap operation.
We give protocols for implementing controlled unitary operations with zero ebits of entanglement resource in Sec.~\ref{SecCUnitary}, controlled traceless operations with one ebit in Sec.~\ref{SecCTLUnitary}, and arbitrary Clifford operations with two ebits in Sec.~\ref{SecClifford}.  In Sec.~\ref{SecBoundCUnitary} we prove that the butterfly network alone can create an entangled state with a Schmidt number of at most 2, and so any operation other than those locally unitary equivalent to a controlled unitary requires a nonzero entanglement resource for implementation.   We show that our protocol is optimal in terms of resource usage in Sec. \ref{Optimality}.  Sec.s \ref{SecCUnitary}, \ref{SecBoundCUnitary} and \ref{Optimality} respectively prove sufficient conditions, necessary conditions, and optimality of the protocol as our main results.  In Sec. \ref{Summary}, a summary and discussions are presented.

\section{Implementation of a swap operation}\label{SecHay}

In this section, we introduce our task of quantum computation over a network,
 and review Hayashi's protocol \cite{Hayashi} for 2-pair communication in the
 context of this task, namely, implementation of a swap operation over the
 butterfly network.

We consider qubit Hilbert spaces and denote the computational basis of a qubit
 as $\{ \ket{0}, \ket{1} \}$. We say that a two-qubitunitary operation $U$ is
 implementable over a network, if we can obtain a joint output state
$U \ket{\psi}$ of qubits at the nodes $B_1$ and $B_2$
 for any input state $\ket{\psi}$ of two qubits, one at the node $A_1$ and another at the node $A_2$, 
 by performing general operations including measurements at each node and communicating qubit and bit information through
 channels specified by edges.   In this paper, we mainly investigate the case where the input state is separable and denoted by $\ket{\psi}=\ket{\psi_1} \otimes \ket{\psi_2}$.  Trivially, if the unitary operation is a tensor product of local unitary operations, it is implementable over any network. 

In Hayashi's protocol \cite{Hayashi} for 2-pair communication,  a special
 butterfly network is described by the nodes $A_1$, $A_2$, $B_1$, $B_2$, $C_1$ and
 $C_2$, and edges $D_1$, $D_2$, $E_1$, $E_2$, $F$, $G_1$, and $G_2$ shown in
 Fig.~\ref{butterfly}.  An additional entanglement resource of 2 ebits is
 shared between the nodes $A_1$ and $A_2$.    The defining characteristic of the
 butterfly network in Hayashi's protocol is that each edge can be chosen to be
 a single-use one way channel with either one qubit quantum capacity or two
 bit classical capacity.  Although a quantum channel of single-qubit capacity
 can send a single-bit of classical information, it cannot faithfully send two
 bits of information.  On the other hand, a classical channel cannot faithfully
 send single-qubit information either.  Thus, the single-qubit quantum and
 2-bit classical channels are mutually inequivalent resources.  Note that
 superdense coding \cite{Superdensecoding} implies that a single-qubit quantum
 channel and shared 1-ebit entanglement together have the capacity of 2-bit
 classical channel, and teleportation shows that a 2-bit classical channel and shared
 1-ebit entanglement together have the capacity of a single-qubit quantum
 channel, however here those ebit resources are not available.

\begin{figure}
\includegraphics[width=7cm]{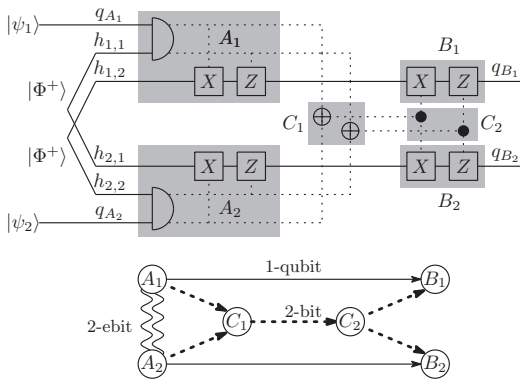}
\caption{\label{hayashi} {\it Upper figure:} The quantum circuit for
 implementing a swap operation on the first qubit and the sixth qubit.
 Each shaded block indicates operations at a node.   $H$ denotes a
 Hadamard operation, and detectors denote Bell measurements in the
 computational basis.  The dotted line represents a controlled operation
 depending on the measurement outcome.  {\it Lower figure:} The butterfly network corresponding to the quantum circuit above,
 showing the amount of communication required in the protocol. The solid line
 denotes a single-qubit channel, and the thin dotted line a single-bit
 channel.}
\end{figure}

The 2-pair communication can be regarded as performing a distributed swap
 operation over the butterfly network, where two arbitrary quantum inputs
 $\ket{\psi_1}$ and $\ket{\psi_2}$ at the nodes $A_1$ and $A_2$, respectively, are
 transferred to the nodes $B_2$ and $B_1$, respectively.  By denoting the input
 qubits at the nodes $A_1$ and $A_2$ by $q_{A_1}$ and $q_{A_2}$, and the output qubits at the node $B_1$ and $B_2$ by $q_{B_1}$ and $q_{B_2}$, respectively, we  can write this as a distributed computation
$
 U_{swap} \ket{\psi_1}_{q_{A_1}} \ket{\psi_2}_{q_{A_2}}=\ket{\psi_2}_{q_{B_1}}  \ket{\psi_1}_{q_{B_2}}.
$
We denote the qubits of the shared ebits at the node $A_1$ by $h_{1,1}$ and
 $h_{1,2}$, while those at the node $A_2$ by $h_{2,1}$ and $h_{2,2}$, see Fig.~\ref{hayashi}.
The qubits ${h}_{i,1}$ and ${h}_{i,2}$ for $i=1,2$ are both in the maximally
 entangled two-qubit state
 \begin{equation}
 \ket{\Phi^+} = ( \ket{00}  +\ket{11} ) / \sqrt{2}.
 \label{maxent}
 \end{equation}
For this protocol, channels $E_1$ and $E_2$ are one qubit quantum channels, while all others are two-bit classical channels.

The protocol is as follows:
\begin{enumerate}
\item At the node $A_1$,  perform a Bell measurement on input qubit
 ${q_{A_1}}$ and  ${h}_{1,1}$  while at the node $A_2$, perform a Bell measurement on the other input qubit ${q_{A_2}}$ and  ${h}_{2,2}$. Let $i, j$ be the two bits of classical information given by the
measurement result at $A_1$ and $k,l$ as that at $A_2$.
Now $X^i Z^j$ and $X^kZ^l$ correspond to the combination of Pauli $X$
and $Z$ corrections for quantum teleportation~\cite{teleportation}
associated with each measurement.
\item At $A_1$, apply $X^iZ^j$ to ${h}_{1,2}$ while at $A_2$, apply $X^kZ^l$ to
      ${h}_{2,1}$.
\item Send qubit ${h}_{1,2}$ from $A_1$ to $B_1$ through the quantum side channel
 $E_1$ and qubit ${h}_{2,1}$ from $A_2$ to $B_2$ through the quantum side channel
 $E_2$.  Send $i, j$ from $A_1$ to $C_1$ and $k,l$ from $A_2$ to $C_1$ via the
 two-bit classical channels $D_1$ and $D_2$ respectively.
\item At $C_1$, compute $i+k, j+l$ (mod $2$).  Then send $i+k, j+l$ to the
node $C_2$ via the two-bit classical channel $F$.
\item Distribute $i+k, j+l$ from $C_2$ to $B_1$ and $B_2$ via the two-bit
 classical channels $G_1$ and $G_2$, respectively.
\item At the node $B_1$, apply the Pauli corrections $X^{i+k}Z^{j+l}$ on
the qubit received from $A_1$ and rename the qubit $q_{B_1}$, and at $B_2$ apply the same operation on the
 qubit received from $A_2$ and rename the qubit $q_{B_2}$.
\end{enumerate}

This protocol can be presented by the quantum circuit and the butterfly network shown in Fig. \ref{hayashi}.  In this circuit, the half circles denote detectors performing Bell measurements described by a set of projectors $\{ \ket{\Phi^{i,j}} \bra{\Phi^{i,j}} \}_{i,j}$ where $\ket{\Phi^{i,j}}= Z^j X^i \ket{\Phi^+}$, the square boxes denote single-qubit operations specified by the letters in the boxes,  and the dotted line represents a controlled operation depending on the measurement outcome.  Note that the symbol $\oplus$ at the node $C_1$ denotes addition of the measurement outcomes modulo 2, it does not represent a controlled operation with classical information at the node $C_2$.  On the other hand, the black circles at the node $C_2$  denote classical control bits for performing the Pauli operations at the nodes $B_1$ and $B_2$.  

In \cite{Hayashi}, it has been shown that this protocol is optimal even for
 asymptotic cases, and that two ebits of entanglement are necessary and
 sufficient for implementing the swap operation (namely, a 2-pair communication),
 in this butterfly network scenario using information theoretical arguments.
   The swap operation is significant since it is the most ``global'' operation in
 terms of entangling power \cite{entanglingpower} and delocalization power
 \cite{delocalizationpower}, compared to controlled unitary operations.
   Our work is motivated by the question of whether or not we can reduce the
 resource requirement by weakening the entangling and delocalization power of the
 network-implemented unitary operations.

\section{Implementation of controlled unitary operations} \label{SecCUnitary}

We consider the deterministic implementation of controlled unitary operations
 over the butterfly network in the setting of Hayashi's protocol,  where we can
 choose a single-qubit quantum channel or a 2-bit classical channel for each
 edge of the network.    We denote a controlled unitary operation by
\begin{equation}
C_{u} = \ket{0}\bra{0} \otimes {\mathbb I}+ \ket{1}\bra{1} \otimes {u}
\end{equation}
 where $u$  is a single-qubit unitary operation.  The controlled unitary operations have
 at most half of the entangling power of the swap operation (which is
 2 ebits), and accordingly they require only half of the resource ebits in
 entanglement-assisted local operations and classical communications (LOCC), where similarly, swap requires 2 ebits. Considering this comparison, it is natural to expect controlled unitary
 operations to require 1 ebit of entanglement shared between the two input
 nodes in order to be implemented over the butterfly network.
   However, we discover a protocol implementing any controlled unitary
 operation over the butterfly network {\it without} using any entanglement
 resource.

\begin{figure}[ht]
\includegraphics[width=5.5cm]{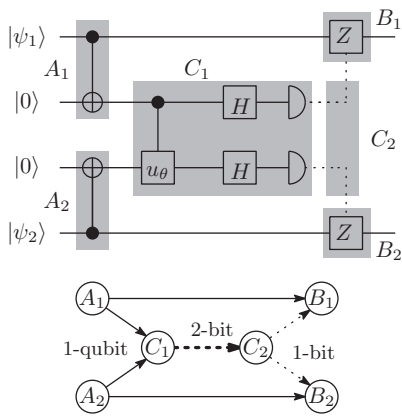}
\caption{\label{cuwoent} {\it Upper figure:} The quantum circuit for
implementing a controlled phase operation on the first qubit and the fourth
 qubit. Each shaded block indicates operations at a node.
   $H$ denotes a Hadamard operation, and detectors denote projective
 measurements in the computational basis ($Z$ measurement).
   The dotted line represents a controlled operation depending on the
 measurement outcome.
{\it Lower figure:} The butterfly network corresponding to the
 quantum circuit above, showing the amount of communication required in the
 protocol. The solid line denotes a single-qubit channel, the thick dotted line
 denotes a two-bit channel and the thin dotted line a single-bit channel.}
\end{figure}

This protocol is based on the implementation of a controlled phase operation
 $C_{u_\theta}$, where a single-qubit phase operation $u_\theta$ is given by
\begin{equation} \label{CPhase}
 u_\theta = \ket{0} \bra{0} + \rm{e}^{i \theta} \ket{1} \bra{1}
 \end{equation}
using the quantum circuit shown in the upper figure of Fig.~\ref{cuwoent}.
 In order to perform $C_{u_\theta}$ over the butterfly network, operations
 shown in each shaded block are performed at each node in the upper figure of
 of Fig.~\ref{cuwoent}, and quantum (classical) information is transmitted
 between the nodes using the quantum (classical) communication specified by the
 edges shown in the lower figure of Fig.~\ref{cuwoent}.

Any controlled unitary operation is locally unitary equivalent to a
 controlled phase operation, namely, we can write
 $C_{u}=({v'}_1 \otimes {v'}_2) C_{{u}_\theta} ({v}_1 \otimes {v}_2)$ using appropriate
 single-qubit unitary operations ${v}_1$, ${v}_2$, ${v'}_1$ and ${v'}_2$.
The protocol implementing $C_{{u}_\theta}$ over the butterfly network can be converted to one implementing $C_u$, where $v_1$ and $v_2$ are first applied on the input by $A_1$ and $A_2$, respectively, then the protocol for $C_{u_\theta}$ is applied, and finally ${v'}_1$ and ${v'}_2$ at the nodes $B_1$ and $B_2$, respectively.
In Sec. \ref{SecBoundCUnitary}, we also show necessity, namely that
 {\it only} controlled unitary operations (and their locally unitary equivalents) are implementable
 over Hayashi's butterfly network setting without using additional entanglement resources.

Note that this protocol does not use the full capacity of the butterfly network
 at the edges $G_1$ and $G_2$, they are only used for transmitting 1 bit,
 instead of the 2-bit capacity allowed.  This extra 1-bit
 capacity could be used for another task, e.g., distributing a shared
 random bit.  It should be also noted that the operations required at nodes
 $A_1$, $A_2$, $B_1$ and $B_2$ do not depend on the angle $\theta$ of the
 controlled phase operation $C_{{u}_\theta}$.  Thus, the distributed quantum
 computation $C_{{u}_\theta}$ can be implemented without revealing the identity
 of the operation to the parties at the input and output nodes.

\section{Implementation of controlled traceless unitary operations} \label{SecCTLUnitary}

In this section, we consider a situation where one of the inner channels, say
 $D_2$, is restricted to a {\it single}-bit classical channel.  We find a
 protocol that implements a slightly weaker class of controlled unitary
 operations, controlled {\it traceless} unitaries, over such a restricted
 butterfly network by adding 1 ebit of entanglement shared between the input
 nodes $A_1$ and $A_2$.   At first sight this protocol consumes more resources
 than the protocol presented in the previous section for implementing a weaker
 class of controlled unitary operations, but as it only requires classical
 communication of 1 bit for the channel $D_2$, comparison of the resource
 requirements between these two protocols is not trivial.

\begin{figure}
\includegraphics[width=4.5cm]{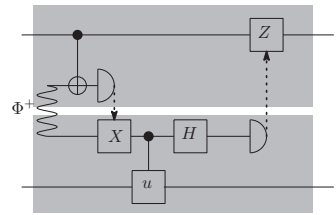}
\caption{\label{LOCCcircuit}The quantum circuit for entanglement-assisted LOCC
 implementation of a controlled unitary operation $C_u$ presented in \cite{LOCC}.
   There are only two nodes; the first two qubits are at the first node (upper
 shaded area) while the third and forth qubits are at the second node (lower shaded area).}
\end{figure}

This protocol is inspired by the entanglement-assisted LOCC implementation of
 controlled unitary operations $C_u$ \cite{LOCC} shown in Fig.~\ref{LOCCcircuit}.
   This LOCC implementation requires a 1-ebit entanglement resource and two-way
 classical communication  (1-bit each way) between the two distant parties.

However, this LOCC implementation is not directly implementable over the
 butterfly network, because in the latter the classical communication is also
 restricted.  This incompatibility is shown in the following way, where a
 similar argument holds for any node at which the controlled unitary operation
 $C_{u}$ appearing in the quantum circuit is performed; here we will assume that
 $C_{u}$ is performed at the node $A_2$.  Since no incoming communication from
 other nodes is allowed at node $A_2$, the classically controlled-$X$ operation
 on the third qubit should also be performed by $A_2$.  Then, the first
 controlled-NOT operation must also be performed at $A_2$ from the same reason.
   But for implementation over the butterfly network, the first qubit should be
 given at node $A_1$ by definition, therefore implementation of a general $C_u$ based on this LOCC implementation scheme is not possible.

Our idea is,  by restricting the class of unitary operations
 $u$,  to find an alternative quantum circuit to implement $C_{u}$ on
 the first and the fourth qubits, where $C_{u}$ on the third and fourth qubits is
 performed at the node $A_2$  before performing any other controlled operation
 required on the third qubit.  If the order of $C_{u}$ on the third and forth qubit and the (classically
 controlled) $X$ operation on the third qubit are changed such that
\begin{equation}
C_{u} (X \otimes {\mathbb I}) = (A \otimes B) C_{u},
\label{condition}
\end{equation}
where $A$ and $B$ are some single-qubit unitary operations to compensate,
 then we arrive at a quantum circuit implementing $C_{u}$ with the desired
 property.

We show that Eq.~(\ref{condition}) is satisfied if and only if the unitary operation $u$ is given by a traceless unitary operation (and its locally unitary equivalents).  To show sufficiency, it is easy to see that for a controlled-$Z$ operation $C_{u}=C_Z$, Eq. (\ref{condition}) is satisfied by taking $A=X$ and $B=Z$, namely, $C_Z \left ( X \otimes {\mathbb I}  \right) = \left ( X \otimes Z \right) C_Z$.  Since any controlled traceless unitary operation $C_{{u}_{tl}}$ can be written as
\begin{eqnarray}
C_{{u}_{tl}}&=&\ket{00} \bra{00}+\ket{01} \bra{01}+e^{i \theta} \ket{10} \bra{10} - e^{i \theta} \ket{11} \bra{11} \nonumber \\
&=& {u}_\theta \otimes {\mathbb I} \cdot C_Z,
\end{eqnarray}
by taking an appropriate basis and using a single-qubit phase operation $u_\theta$ defined in Eq.~(\ref{CPhase}), one can implement any controlled traceless operation.

To show necessity, we first rearrange Eq.~(\ref{condition}) as
\begin{equation}
A \otimes B = C_{u} (X \otimes \mathbb{I} ) C_{u}^\dagger = \ket{1}\bra{0}\otimes
 {u}+\ket{0}\bra{1}\otimes {u}^\dagger.
\label{AB}
\end{equation}
By taking partial traces of Eq.~(\ref{AB}), the following two conditions
\begin{eqnarray}
\mathrm{Tr}_A (A \otimes B) = 0
\label{trA}
\end{eqnarray}
and
\begin{eqnarray}
\mathrm{Tr}_B (A \otimes B)=(\Re\,\mathrm{Tr}_B {u})X+(\Im\,\mathrm{Tr}_B {u}) Y
\label{trB}
\end{eqnarray}
have to be satisfied.   For Eq.~(\ref{trA}), the case of $B=0$ is uninteresting, so we consider the case given by $\mathrm{Tr}_A A = 0$ and denote $A$'s
eigenvalues by $\pm \alpha$.  Then $B$'s eigenvalues are $\pm 1/\alpha$ or both $1/\alpha$, since the eigenvalues of $A \otimes B$ are equal to those of $X\otimes\mathbb{I}$, which are $\pm 1$.  The case when $B$'s eigenvalues are degenerate is trivial, $B$ is equal to the identity up to some factor.  Otherwise $\mathrm{Tr}_B B=0$ and from Eq.~(\ref{trB}) we can conclude
$\mathrm{Tr} {u}=0$.  Thus, only controlled traceless unitary operations can satisfy Eq.~(\ref{condition}).

\begin{figure}
\includegraphics[width=6cm]{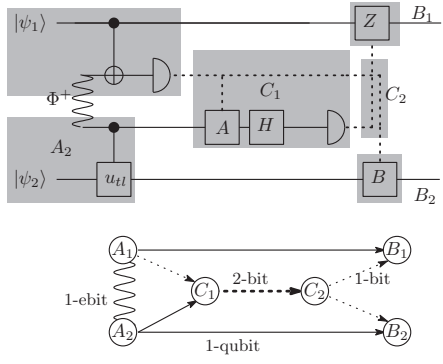}
\caption{\label{tr0p1}
{\it Upper figure:} The quantum circuit for implementing a controlled
 traceless unitary operation $C_{{u}_{tl}}$ on the first qubit and the fourth qubit.
{\it Lower figure:} The butterfly network corresponding to
 the quantum circuit above, showing the amount of communication required in
the protocol.}
\end{figure}

The corresponding quantum circuit for this implementation of a controlled-traceless operation over the butterfly network is shown in the upper part of Fig.\ref{tr0p1}.
By performing the operations given in each shaded block at each node,
and transmission of quantum or classical information between the nodes specified
by the edges shown in the lower figure of Fig.\ref{tr0p1}, a controlled traceless unitary operation
$C_{{u}_{tl}}$ is implementable over the butterfly network.

\section{Implementation of Clifford operations}\label{SecClifford}

In this section, we construct a protocol for implementing Clifford operations on
 the butterfly network by slightly modifying the protocol for the swap
 operation $U_{swap}$ of Sec. \ref{SecHay}.  Here, a Clifford operation
 $U_{Cl}$ is defined as any operation that maps the Pauli group to itself, the
 group of which is known to be generated by a controlled-NOT operation, a
 Hadamard operation $H$, a phase operation $S=\ket{0}\bra{0}+i \ket{1}\bra{1}$,
 and Pauli operations.
Any two-qubitClifford operation can be written in the form of
 $U_{Cl} \cdot U_{swap}$ by an appropriate choice of $U_{Cl}$, since $U_{swap}$
 also belongs to the Clifford group.   Here we construct a protocol for
 implementing $U_{Cl} \cdot U_{swap}$ over the butterfly network.

Suppose that a given Clifford operation $U_{Cl}$ satisfies $U_{Cl} (X_1
\otimes \mathbb{I})=(P_1\otimes P_2) U_{Cl}$ and $U_{Cl}(Z_1\otimes
\mathbb{I})=(Q_1\otimes Q_2) U_{Cl}$, where $P_1$, $P_2$, $Q_1$, $Q_2$
 represent Pauli operators.  The initial state of the protocol is given by
$\ket{\psi_1}_{q_{A_1}}\ket{\Phi^+}_{{h}_{1,1} {h}_{2,1}} \ket{\psi_2}_{q_{A_2}}
\ket{\Phi^+}_{{h}_{2,2} {h}_{1,2}}$, using the notation introduced in Sec.
 \ref{SecHay}.

First, perform a Bell measurement on
${q_{A_1}}$ and ${h}_{1,1}$ at the node $A_1$ and then perform $U_{Cl}$ at the
node $A_2$ on ${h}_{2,1}$ and ${q_{A_2}}$.  By denoting the measurement outcomes at the
 node $A_1$ to be $i,j$,  the resulting state can be written as
\begin{eqnarray}
\ket{{\tilde \Phi}^{i j}}_{{q_{A_1} h}_{1,1}}
 \ket{{\tilde \psi}_{1 2}^{i j}}_{{h}_{2,1} {q_{A_2}}} \ket{\Phi^+}_{{h}_{2,2} {h}_{1,2}},
\end{eqnarray}
where the states
\begin{equation}
\ket{{\tilde \Phi}^{i j}}=(X^i Z^j\otimes\mathbb{I}) \ket{\Phi^+}
\end{equation}
and
\begin{equation}
\ket{{\tilde \psi}_{12}^{i j}}=(P_1\otimes P_2)^i(Q_1\otimes Q_2)^j U_{Cl}
 \ket{\psi_1}  \ket{\psi_2}
\end{equation}
denote the post measurement states corresponding to the outcome $i,j$.

Next, perform a second Bell measurement on ${q_{A_2}}$ and ${h}_{2,2}$ at the
node $A_2$ and denote the measurement outcomes by $k,l$.  This effects a teleportation of $\ket{\psi_2}$.  The state is now transformed to
\begin{eqnarray}
\ket{{\tilde \Phi}^{i j}}_{{q_{A_1}}  {h}_{1,1}} \ket{{\tilde \psi}_{12}^{i j k l}}_{{h}_{1,2} {h}_{2,1}} \ket{{\tilde \Phi}^{k l}}_{q_{A_2}  {h}_{2,2}}
\end{eqnarray}
where
\begin{equation}
\ket{{\tilde \psi}_{12}^{i j k l}} = (\mathbb{I} \otimes X_2^k Z_2^l)
 (P_1 \otimes P_2)^i(Q_1\otimes  Q_2)^j U_{Cl}  U_{swap} \ket{\psi_1}
 \ket{\psi_2}
\end{equation}
denotes the post measurement state after the second Bell measurement,
 corresponding to the outcome $i,j,k,l$.

The parties at the nodes $A_1$ and $A_2$ now hold the uncorrected outputs
 ${h}_{1,2}$ and ${h}_{2,1}$, respectively. $A_1$ then performs $X_2^i Z_2^j P_2^i Q_2^j$ on the qubit $h_{1,2}$ while $A_2$ performs $P_1^k Q_1^l$ on ${h}_{2,1}$, sending their measurement outcomes
 to the node $C_2$, just as in the protocol in \cite{Hayashi}.

The parties at the $B$ nodes receive the classical outcomes $i+k$ and $j+l$ from
 the corresponding $A$ nodes.
The party at node $B_1$ can correct the state  by performing
 $X_1^{i+k}Z_1^{j+l}$ on her received qubit and renames it ${q_{B_1}}$, whereas at node $B_2$, the party performs
 $P_1^{i+k}Q_1^{j+l}$ and renames the qubit $q_{B_2}$.  This completes the protocol.

\section{Necessity of controlled unitary operations}
\label{SecBoundCUnitary}

In Sec. \ref{SecCUnitary}, we showed that global unitary operations are implementable
over the butterfly network in Hayashi's setting without adding any extra resources, if
they are controlled-unitary operations.   In this section, we show
that global unitary operations are implementable over the butterfly network without
adding any extra resources {\it only if} they are controlled unitary operations.

To prove this, we consider the entangling capability \cite{comment} of the butterfly network for creating entangled states at the output nodes $B_1$ and $B_2$, for a separable input state given at the nodes $A_1$ and $A_2$.  We analyze this capability in terms of the Schmidt numbers, the number of non-zero coefficients in the Schmidt decomposition of a bipartite entangled pure state, of the output states.

\begin{figure} 
\includegraphics[width=6cm]{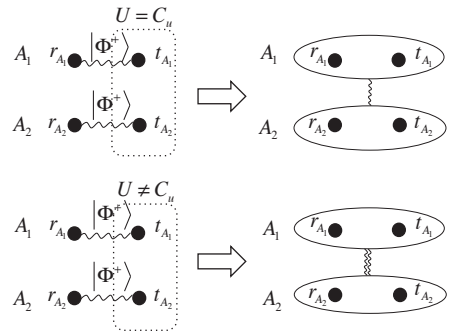}
\caption{Comparison of the capability of 2-qubit global unitary operations for creating entangled states.  The unitary operation
is applied to the qubits $t_{A_1}$ and $t_{A_2}$ at the nodes $A_1$ and $A_2$, respectively, each of which is entangled with another qubit ${r}_{A_1}$ (at the node $A_1$) and ${r}_{A_2}$ (at the node $A_2$).    A controlled unitary operation $C_u$ can create an entangled state with Schmidt number only up to 2 (depicted by a single wavy line), and a global unitary operation $U \neq C_u$ creates one with Schmidt number 4 (double wavy line).}
\label{schmidtnumber}
\end{figure}

For a bipartite unitary operation $U$, the entangling capability can be evaluated by examining a four-qubit entangled state obtained by applying $U$ to two qubits each of which is maximally entangled with another qubit, (see Fig. \ref{schmidtnumber}).  By denoting the two qubits to which $U$ is applied by $t_{A_1}$ and $t_{A_2}$, and the corresponding maximally entangled qubits by $r_{A_1}$ and $r_{A_2}$, the four-qubit state is given by
\[
U \otimes \mathbb{I}_{r_{A_1},r_{A_2}}  \ket{\Phi^+}_{t_{A_1},r_{A_1}}  \ket{\Phi^+}_{t_{A_2},r_{A_2}} ,
\]
where $\ket{\Phi^+}_{t_{A_k},r_{A_k}}$ denotes a maximally entangled state of the qubits $t_{A_k}$ and $r_{A_k}$ defined by Eq.(\ref{maxent}) for $k=1,2$.  In \cite{SchmidtDecomp}, it is shown that any controlled unitary operation $U=C_u$ can create an entangled state with Schmidt number only up to 2, and also that any global operation that is not locally unitary equivalent to a controlled unitary operation, which we denote by $U \neq C_u$, must create an entangled state with Schmidt number 4.

By applying this result to the unitary operation implemented by the butterfly network, we can say that if $U \neq C_u$ can be deterministically implemented over the butterfly network (without adding resources) for a pure input state $\ket{\Phi^+}_{t_{A_1},r_{A_1}} \ket{\Phi^+}_{t_{A_2},r_{A_2}} $, then we can deterministically create an entangled state shared among the nodes $A_1$, $A_2$, $B_1$ and $B_2$ with Schmidt number 4 in terms of the bipartite partition of nodes $\{ \{ A_1, B_1 \} , \{ A_2, B_2 \} \}$.

Taking the contrapositive, the statement is that if a pure entangled state shared between the nodes $A_1$, $A_2$, $B_1$ and $B_2$ with Schmidt number 4 in terms of the bipartite partition $\{ \{ A_1, B_1 \} , \{ A_2, B_2 \} \}$ cannot be deterministically created by using the butterfly network with a pure input state given by $\ket{\Phi^+}_{t_{A_1},r_{A_1}} \ket{\Phi^+}_{t_{A_2},r_{A_2}} $, then $U \neq C_u$ {\it cannot} be deterministically implemented over the butterfly network.  Thus, what we need to show is the impossibility of deterministically creating an entangled pure state with Schmidt number 4 using the butterfly network.

\begin{figure} 
\includegraphics[width=3.5cm]{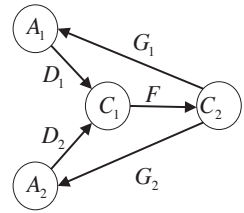}
\caption{A ``collapsed" butterfly network obtained from Fig. \ref{butterfly} by identifying $B_1$ with $A_1$ and $B_2$ with $A_2$, and removing the outer channels $E_1$ and $E_2$.   Each arrow can be chosen to be a 1-qubit quantum channel or a 2-bit classical channel in this setting.}
\label{stg_butterfly}
\end{figure}

To do so, we investigate the entangling capability of the butterfly network using the ``collapsed'' butterfly network shown by Fig. \ref{stg_butterfly}.  This collapsed butterfly network is obtained by identifying the nodes $B_1$ with $A_1$ and the nodes $B_2$ with $A_2$, and removing the outer channels $E_1$ and $E_2$ of the original butterfly network.   In this section, we allow the inclusion of (untransmitted) ancilla qudits (quantum $d$-level systems) in order to facilitate general operations at any node.  The channels represented by the remaining edges $D_1$, $D_2$, $F$, $G_1$, $G_2$ of the collapsed butterfly network can be chosen to be either 1-qubit  quantum channels or 2-bit classical channel, following Hayashi's setting.

The collapsed butterfly network can be viewed as the butterfly network with additional resources, namely, free undirected quantum and classical communication between $A_1$ and $B_1$, and also between $A_2$ and $B_2$.   In the following, we prove by contradiction
that even when we use this ``stronger" network, it is impossible to deterministically create a pure bipartite entangled state shared between $A_1$ and $A_2$ with Schmidt number 4, given that the initial state is prepared by a tensor product of pure states in each node, and there is no entanglement between the qubits and qudits at different nodes.

In order to arrive at a contradiction, we assume that by using the collapsed butterfly network with the initial state $\ket{\Phi^+_{A_1}} \ket{\Phi^+_{A_2}}$ at the nodes $A_1$ and $A_2$, it is possible to create a final state with Schmidt number 4 between the qubits at the nodes $A_1$ and $A_2$
\begin{eqnarray}
\ket{\Theta}=\sum_{k=0}^3 \sqrt{\lambda_k} \ket{\alpha_k}  \ket{\beta_k},
\label{finalschmidt0}
\end{eqnarray}
where $\lambda_k$ are nonzero Schmidt coefficients satisfying $\sum_k {\lambda_k} = 1$ and $\{ \ket{\alpha_k} \}$ and $\{ \ket{\beta_k} \}$ are orthonormal bases for the two qubits at the nodes $A_1$ and $A_2$, respectively.

A general protocol for converting the initial state $\ket{\Phi^+_{A_1}} \ket{\Phi^+_{A_2}}$ into the final pure entangled state $\ket{\Theta}$ using the collapsed butterfly network can be described by the following steps.
\begin{enumerate}
\item{Performing general operations independently at the nodes $A_1$ and $A_2$.}
\item{Transmission of 1 qubit, denoted $t_{A_1}$, or 2 bits, denoted $c_{A_1}$, from the node $A_1$ to $C_1$ using channel $D_1$, and similarly for $t_{A_2}$ or $c_{A_2}$ from node $A_2$ to $C_1$ along channel $D_2$.}
\item{Performing a general operation at the node $C_1$.}
\item{Transmission of 1 qubit, $t_{C_1}$, or 2 bits, $c_{C_1}$, from the node $C_1$ to $C_2$ using channel $F$.}
\item{Performing a general operation at the node $C_2$.}
\item{Transmission of 1 qubit, $t_{C_2}$, or 2 bits, $c_{C_2}$, from the node $C_2$ to $A_1$ using channel $G_1$, and similarly for $t'_{C_2}$ or $c'_{C_2}$ from the node $C_2$ to $A_2$ along channel $G_2$.}
\item{Performing general operations independently at the nodes $A_1$ and $A_2$.}
\end{enumerate}

\begin{figure}
\includegraphics[width=7.5cm]{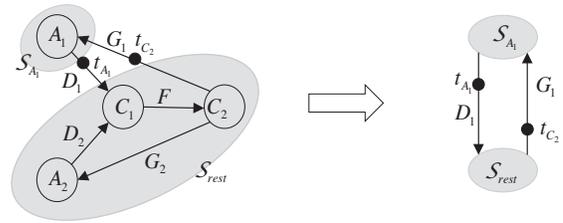}
\caption{The bipartite picture in terms of the partition $\mathcal{S}_{A_1}=\{ A_1 \}$ and $\mathcal{S}_{rest}=\{ C_1, C_2, A_2 \}$ of the collapsed butterfly network. Black circles denote qubits $t_{A_1}$ transmitted from $A_1$ using the channel $D_1$ and $t_{C_2}$ transmitted from $C_2$ using the channel $G_1$.}
\label{strong-butterfly3}
\end{figure}

First, we show that in the steps 2 and 6, both of the channels $D_1$ and $G_1$ should be used as quantum channels, not classical channels.   By grouping the nodes into two sets $\mathcal{S}_{A_1}=\{ A_1 \}$ and $\mathcal{S}_{rest}=\{ C_1, C_2, A_2 \}$, as shown in Fig.  \ref{strong-butterfly3}, it is easy to see that both of the channels $D_1$ and $G_1$ should be quantum in order to create a bipartite entangled state with Schmidt number greater than 2 for this partition.  Since the final state $\ket{\Theta}$ is a special case of a bipartite state with Schmidt number 4 in terms of this partition,  both $D_1$ and $G_1$ should be used as quantum channels.  In a similar manner, by introducing the partition $\{ \mathcal{S}_{A_2}=\{ A_2 \} , \mathcal{S}'_{rest}=\{ C_1, C_2, A_1 \} \}$, we can also see that $D_2$ and $G_2$ should be quantum channels.

This picture also makes it clear that the channels $D_1$, $G_1$, $D_2$ and $G_2$ should be used for transmitting a qubit that is entangled with another qubit (or several qubits/qudits) kept at the same set of the nodes.   Thus,  at step 1, the general operation performed at the node $A_1$ should not disentangle the qubit to be transmitted, $t_{A_1}$, from other qubits and qudits at $A_1$.   Similarly, the qubit  $t_{A_2}$ should not be disentangled from node $A_2$.

In general, after step 1, the state at the node $A_1$ can be a mixed state, that is, it can be a pure state entangled with an extra ancilla qudit $r'_{A_1}$ at the node $A_1$ as well as the qubit $r_{A_1}$, which was initially maximally entangled with ${t}_{A_1}$.   The qudit $r'_{A_1}$ should be considered inaccessible from any node, including $A_1$.  However, because what we want to show in this section is the impossibility of the transformation from $\ket{\Phi^+_{A_1}} \ket{\Phi^+_{A_2}}$ to $\ket{\Theta}$ using the butterfly network, it suffices to show impossibility for the relaxed situation where we regard $r'_{A_1}$ as accessible at $A_1$.

\begin{figure}
\includegraphics[width=5cm]{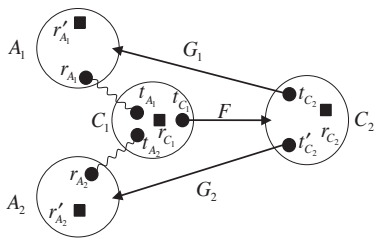}
\caption{The schematic picture of the state just after step 2, having transmitted $t_{A_1}$ and $t_{A_2}$ to $C_1$ using quantum channels $D_1$ and $D_2$.   The black circles represent transmitted qubits $t_{A_1}$ and $t_{A_2}$ and ancilla qubits $r_{A_1}$ and $r_{A_2}$.  Each wavy line represents the existence of entanglement with Schmidt number 2.   The other qubits to be transmitted $t_{C_1}$, $t_{C_2}$ and $t'_{C_2}$ are also represented by black circles, and ancilla qudits $r_{C_1}$, $r_{C_2}$ at the nodes $C_1$, $C_2$ are represented by black squares.   Arrows represents channels not used in this step.}
\label{step2}
\end{figure}

When the state at node $A_1$ after step 1 is given by an entangled state of qubits $t_{A_1}$, $r_{A_1}$ and a qudit $r'_{A_1}$, it is always possible to perform a unitary operation $W_{A_1}$ on $r_{A_1}$ and $r'_{A_1}$ that transforms the state into a tensor product of a two-qubitstate of $t_{A_1}$, $r_{A_1}$ and a qudit state of $r'_{A_1}$.  Since we can compensate for $W_{A_1}$ by performing $W_{A_1}^\dagger$ on $r_{A_1}$ and $r'_{A_1}$ at step 7, we only need to consider the general operations that map a state $\ket{\Phi^+_{A_1}}$ to another pure entangled qubit state.   A similar argument holds for a general operation at the node $A_2$.   In Fig. \ref{step2}, we show a schematic picture of the state just after step 2.

Next, we observe that after step 3, but before step 4, any qubit and qudit at node $C_2$ cannot be entangled with other qubits and qudits at the nodes $A_1$, $A_2$, and $C_1$.   If the channel $F$ is used as a classical channel at step 4, any qubit and qudit at the node $C_2$ remains separable from any other qubits and qudits at the nodes $A_1$, $A_2$, and $C_1$.

\begin{figure}
\includegraphics[width=5cm]{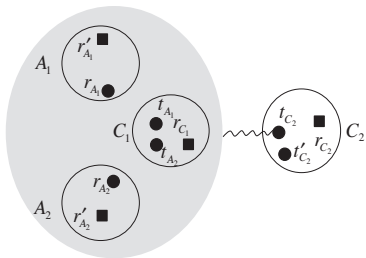}
\caption{The schematic picture of the state just before step 5, after transmitting $t_{C_1}$ to $C_2$ using quantum channel $F$.   Wavy lines represent the existence of entanglement where the rank of the reduced density matrix $\sigma_{t_{C_2}}$ is at most 2.  The shaded region represents a group of qubits and qudits that are entangled with the qubit $t_{C_2}$ and the wavy line represents entanglement with the Schmidt number 2 that gives the rank of the reduced density matrix $\sigma_{t_{C_2}}$  of at most 2.  The qubit $t'_{C_2}$ and $r_{C_2}$ are not entangled with the qubit $t_{C_2}$. In this figure, the unused channels $G_1$ and $G_2$ are not shown.  }
\label{beforestep5}
\end{figure}

If the channel $F$ is used as a quantum channel, a qubit $t_{C_1}$ is transmitted from $C_1$ to $C_2$.   After the transmission, the qubit $t_{C_1}$ is renamed $t_{C_2}$.  Just before step 5, $t_{C_2}$ can be entangled with qubits or qudits located outside node $C_2$, but the rank of the reduced density matrix $\sigma_{t_{C_2}}$ is at most 2 (see Fig. \ref{beforestep5}).

In order to derive conditions on the generalized operations performed at node $C_2$ in step 5, we examine the protocol from the reverse, and investigate the conditions on general operations performed at the nodes $A_1$ and $A_2$ in step 7.
A general operation is described by an \textit{instrument}  \cite{instrument}, which is defined  by a set of completely positive maps, $\{ \Epsilon^{(i)} \}$, such that
 \[
  \mathrm{Tr} \sum_i \Epsilon^{(i)} (\rho) = 1
 \]
for any normalized density matrix $\rho$.  Let us denote the joint state of qubits $t_{C_2}$, $t'_{C_2}$, $r_{A_1}$ and  $r_{A_2}$ obtained just after step 6 by $\tilde{\rho}$.  Note that in this step, $t_{C_2}$ and $r_{A_1}$ are qubits at the node $A_1$, while $t'_{C_2}$ and $r_{A_2}$ are qubits at $A_2$.  We also denote the reduced density matrix of the qubits at the node $A_1$ by $\tilde{\rho}_{A_1} =\mathrm{Tr}_{t'_{C_2},r_{A_2}} \tilde{\rho}$, and that at the node $A_2$ by $\tilde{\rho}_{A_2} =\mathrm{Tr}_{t_{C_2},r_{A_1}} \tilde{\rho}$.  Since these are the last operations in the protocol and no more communication is allowed, the generalized operation should return the final pure state $\ket{\Theta}$ for any outcome $i$,  namely,
\[
 \Epsilon^{(i)} (\tilde{\rho}_{A_k}) = p^{(i)} \Theta_{A_k},
\]
where $p^{(i)}$ satisfies
\[
 \sum_i p^{(i)} = 1,
\]
and $\Theta_{A_k}$ is the reduced density matrix of $\ket{\Theta}$ at the node $A_k$.
This implies that there exists a completely positive trace preserving (CPTP) map that transforms $\tilde{\rho}_{A_k}$ to $\Theta_{A_k}$, namely,
\[
 \Lambda_{A_k} (\tilde{\rho}_{A_k}) = \Theta_{A_k}.
\]
We denote the combination of the CPTP maps at the nodes $A_1$ and $A_2$ by $\Lambda_{A_1} \otimes \Lambda_{A_2}$.

We investigate the conditions for the state $\tilde{\rho}$ to be transformable to the final state $\ket{\Theta}$ by the CPTP map $\Lambda_{A_1} \otimes \Lambda_{A_2}$.  In the Appendix, Lemma 1 shows that for such a transformation to be possible, the state $\tilde{\rho}$ shared between the nodes $A_1$ and $A_2$  has to be a pure entangled state $\ket{\varphi}$ with Schmidt coefficients equivalent to those of $\ket{\Theta}$.   Since step 6 only changes the locations of qubits, the state of qubits $t_{C_2}$, $t'_{C_2}$, $r_{A_1}$ and $r_{A_2}$ just after step 5 should also be given by the pure state $\ket{\Theta}$.

If the generalized operation in step 5 is described by a generalized measurement on the qubits $t_{C_2}$ and $t'_{C_2}$, then the state after the measurement depends on the measurement outcomes.  However, as we have shown, the state after step 5 should be given deterministically by a pure state $\ket{\Theta}$.  Therefore, this operation can also be described by a CPTP map $\Lambda_{C_2}$ on  the qubits $t_{C_2}$ and $t'_{C_2}$.

We denote the density matrix of the qubits $t_{C_2}$, $r_{A_1}$ and $r_{A_2}$ just before step 5 by $\sigma$.  It is sufficient to consider the qubit $t'_{C_2}$ and (sufficiently large) qudit ${r}_{C_2}$ in fixed states both denoted by $\ket{0}$.   Using a Stinespring dilation, the action of the CPTP map  $\Lambda_{C_2}$ can be written
\begin{eqnarray}
\left[ \mathbb{I}_{A_1,A_2} \otimes \Lambda_{C_2} \right] (\sigma \otimes \ket{00} \bra{00})
= \mathrm{Tr}_{ r_{C_2}}  \sigma'
\end{eqnarray}
by using a density matrix $\sigma'$ for the extended system $r_{A_1}$, $r_{A_2}$, $t_{C_2}$, $t'_{C_2}$ and $r_{C_2}$ given by
\begin{eqnarray}
\sigma'= (\mathbb{I}_{A_1, A_2} \otimes V_{C_2}) (\sigma  \otimes \ket{00} \bra{00}) (\mathbb{I}_{A_1, A_2} \otimes V_{C_2}^\dagger)
\end{eqnarray}
where $V_{C_2}$ denotes a unitary operation on the qubits $t_{C_2}$ and $t'_{C_2}$, while $\ket{00}$ denotes the fixed joint state of the qubit  $t'_{C_2}$ and qudit ${r}_{C_2}$.

\begin{figure}
\includegraphics[width=5cm]{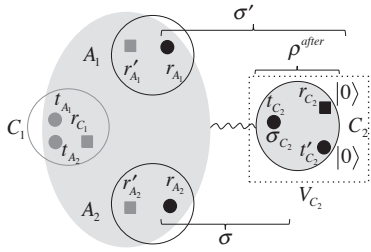}
\caption{Schematic picture of the situation just after step 5 as required by the situation just after step 4.  The CPTP map $\Lambda_{C_2}$ on qubits $t_{C_2}$ and $t'_{C_2}$ is rewritten by adding ancilla ${r}_{C_2}$ and performing a unitary operation $V_{C_2}$.  Thus the two shaded regions form a pure bipartite entangled state with Schmidt number 2,  which is represented by the wavy line, and the rank of $\rho^{\rm after}$ is at most 2.  Systems irrelevant for analyzing the rank of $\rho^{\rm after}$ are depicted in gray.}
\label{afterstep5-forward}
\end{figure}

Using the density matrices $\sigma$ and $\sigma'$, we define the reduced density matrix of the qubits $t_{C_2}$ and $t'_{C_2}$ and the qudit $r_{C_2}$ before step 5 by
\[
\rho^{\rm before} =(\mathrm{Tr}_{r_{A_1}, r_{A_2}}  \sigma) \otimes \ket{00} \bra{00} = \sigma_{t_{C2}} \otimes \ket{00} \bra{00},
\]
and that after step 5 by
\[
\rho^{\rm after} = \mathrm{Tr}_{r_{A_1}, r_{A_2}} \sigma'.
\]
We present a schematic picture of this situation in Fig.~\ref{afterstep5-forward}.

The ranks of $\rho^{\rm before} $ and $\rho^{\rm after}$ should be the same, since the two states are related by a unitary operation $V_{C_2}$ on $t_{C_2}$, $t'_{C_2}$ and $r_{C_2}$.   The rank of $\rho^{\rm before}$ is given by the rank of $\sigma_{t_{C_2}}$, which we have shown to be less than or equal to 2, (see Fig. \ref{beforestep5}).  For the case in which the channel $F$ is used as a quantum channel, therefore, the ranks of $\rho^{\rm before} $ and $\rho^{\rm after} $ are both at most 2, (see Fig. \ref{afterstep5-forward}).  For the case in which the channel $F$ is used as classical channel, we can always prepare a pure state for the qubit $t_{C_2}$, and therefore $\mathrm{rank} (\rho^{\rm after}) = 1$.

On the other hand, from the requirement of the state of the qubits $t_{C_2}$, $t'_{C_2}$, $r_{A_1}$ and $r_{A_2}$ to be pure, we obtain the relationship
\begin{eqnarray}
\sigma'= \ket{{\Theta}} \bra{{\Theta}} \otimes \sigma'_{r_{C_2}}
\end{eqnarray}
where $\sigma'_{r_{C_2}}$ is a state of qudit $r_{C_2}$.   From this relation, the rank of $\rho^{\rm {after}}$  is given by
\[
\mathrm{rank} (\rho^{\rm after}) = \mathrm{rank} (\mathrm{Tr}_{r_{A_1}, r_{A_2}} \ket{\Theta} \bra{\Theta})  \cdot  \mathrm{rank} (\sigma'_{r_{C_2}}).
\]

\begin{figure}
\includegraphics[width=5cm]{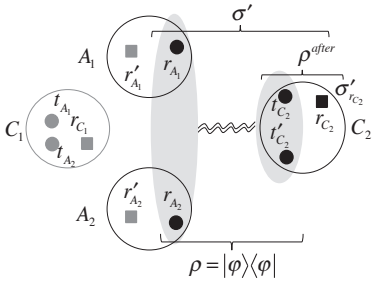}
\caption{Schematic picture of the situation just after step 5 as required by the situation just before step 6, that is, looking at the protocol in the reverse direction.  The two shaded regions form a pure bipartite entangled state with Schmidt number 4, which is represented by the double wavy line, and the rank of $\rho^{\rm after}$ is 4.  Systems irrelevant for analyzing the rank of $\rho^{\rm after}$ are depicted in gray.}
\label{afterstep5-backword}
\end{figure}

The rank of the reduced density matrix of the qubits $r_{A_1}$ and $r_{A_2}$ just after step 2 is 4, since each of the qubits is entangled with another qubit.  After step 2, the rank of this reduced density matrix remains the same until just before step 7, since no operation is applied to the qubits $r_{A_1}$ and $r_{A_2}$. After step 5, the 4-qubit state of $t_{C_2}$, $t'_{C_2}$, $r_{A_1}$ and $r_{A_2}$ should be the pure state $\ket{\Theta}$, and therefore the rank of the reduced density matrix of qubits $t_{C_2}$ and $t'_{C_2}$ is also 4: $\mathrm{rank} (\mathrm{Tr}_{r_{A_1}, r_{A_2}} \ket{\Theta} \bra{\Theta}) = 4$,  (see Fig. \ref{afterstep5-backword}). Since the Schmidt rank of entangled states cannot be changed by local operations (even
probabilistically and with classical communications \cite{SLOCC}), the relationship
\[
k \geq \mathrm{rank} (\rho^{\rm before}) = \mathrm{rank} (\rho^{\rm after}) = 4 \cdot \mathrm{rank} (\sigma'_{r_{C_2}})
\]
should be satisfied, where $k=1$ for the case when $F$ is used as a classical channel, and $k=2$ for the case when it is quantum.   Since $\mathrm{rank} (\sigma'_{r_{C_2}}) \geq 1$, this relation cannot be satisfied, and our contradiction has been reached.

\section{Optimality of our protocol for controlled unitary operations} \label{Optimality}

In Sec. \ref{SecCUnitary}, we presented a protocol for implementing controlled unitary operations over the butterfly network where the four channels $D_1$, $D_2$, $E_1$, and $E_2$ are used as 1-qubit quantum channels, $F$ is a 2-bit classical channel, and $G_1$, $G_2$ are 1-bit classical channels, with no consumption of entanglement.  In this section, we show that controlled unitary operations cannot be implemented over the butterfly network if we change the specifications of any of these channels, and therefore that the protocol is optimal in terms of resource usage.

First, we prove that the use of four 1-qubit quantum channels is necessary.  We consider the situation where the inputs at the node $A_1$ and $A_2$ are parts of maximally entangled states within the nodes, namely, the input state is given by $\ket{\Phi^+}_{t_{A_1},r_{A_1}} \ket{\Phi^+}_{t_{A_2},r_{A_2}}$.  By denoting the qubits representing the outputs by $r'_{B_1}$ and $r'_{B_2}$, the final state of the protocol for implementing $U$ over the butterfly network is given by
$$
\left ( U \otimes \mathbb{I}_{r_{A_1},r_{A_2}} \right ) \ket{\Phi^+}_{r'_{B_1},r_{A_1}} \ket{\Phi^+}_{r'_{B_2},r_{A_2}}.
$$
This state has Schmidt number 4 between the partition of the nodes $\{ \{ {A_1}, {A_2} \} , \{ B_1, B_2 \} \}$.  The Schmidt number between the partition of the nodes $\{ \{ A_1, B_1 \} , \{ A_2, B_2 \} \}$ is 2 for the case of $U=C_u$, as we have shown in the previous section.

To create a pure entangled state with Schmidt number $4$ between the bipartite partition from a tensor product of pure states at the input nodes using a quantum network, at least two quantum channels should connect the set including the nodes $A_1$ and $A_2$ and the set including $B_1$ and $B_2$.  To create an entangled state with Schmidt number $2$, at least one quantum channel should connect the set including the nodes $A_1$ and $B_1$ and the set including the nodes $A_2$ and $B_2$. The final state of nodes $C_1$ and $C_2$ are arbitrary, so we take the state requiring the least amount of resource, namely, a pure product state.   Then the nodes $C_1$ and $C_2$ can be included in any bipartite partition we like.  The condition for the number of connections between the sets of nodes required by the Schmidt number should be satisfied for all of these inclusions of $C_1$ and $C_2$.   It is not hard to see that if we use only three quantum channels, a final state with the required Schmidt number cannot be achieved.  Thus, at least four quantum channels are necessary.

Second, we show that the four quantum channels should be $D_1$, $D_2$, $E_1$, and $E_2$.   It is also straightforward to see that each input and output node $A_1$, $A_2$, $B_1$, and $B_2$ should be connected to at least one quantum channel, otherwise we cannot maintain entanglement of the final state.   If $E_1$ is not chosen as a quantum channel, then $D_1$ and $G_1$ should be quantum channels.  However, in this case, we can never achieve the final state no matter how we assign the other two quantum channels.  Therefore, $E_1$ should be used as a quantum channel and similarly, so should $E_2$.  Assignment of the remaining two quantum channels is thus narrowed down to the pairs $\{ D_1, D_2 \}$ or $\{ G_1, G_2 \}$.

We can exclude the pairs $\{ G_1, G_2 \}$ by the following argument. If we choose $G_1$ and $G_2$ to be quantum channels $D_1$ and $D_2$ must be used as classical channels in order to not exceed our resource limit. To satisfy the Schmidt number requirement,  one of the qubits in an entangled state $\ket{\Psi}$ at $C_2$ should be sent by $G_1$ and the other by $G_2$.  Now we consider two arbitrary inputs, $\ket{\psi_1}$ at node $A_1$ and $\ket{\psi_2}$ at $A_2$, given for the collapsed butterfly network introduced in Sec.~\ref{SecBoundCUnitary}.  If $C_u$ can be implemented over the original butterfly network, it should also implementable over the collapsed butterfly network. Implementation over the collapsed network can be regarded as LOCC implementation of $C_u$ assisted by entanglement given by $\ket{\Psi}$.  However, entanglement-assisted deterministic LOCC implementation requires two-way communication between the nodes \cite{LOCCimplementation}, therefore, implementation of $C_u$ is not possible if we use $G_1$ and $G_2$ as quantum channels.  Thus, $D_1$ and $D_2$ should be used as quantum channels in addition to $E_1$ and $E_2$.

Third, we show that the channel $F$ should be a 2-bit classical channel and cannot be a 1-bit classical channel.  We show this by contradiction.   Assume that only 1 bit of classical communication is necessary over the channel $F$.  Then it is also possible to implement $C_u$ over the collapsed butterfly network.   Because the channels  $F$, $G_1$ and $G_2$ are now all at most 1-bit classical channels, and the final state is an entangled state between the nodes $A_1$ and $A_2$ in general, the channels $D_1$ and $D_2$ in Fig. \ref{stg_butterfly} cannot be used to send separable states --- the transmitted qubits $t_{A_1}$ and $t_{A_2}$ are parts of entangled states, which can be written as $\ket{\Psi_{1}}_{ t_{A_1},r_{A_1},r'_{A_1}}$ and $\ket{\Psi_{2}}_ {t_{A_2},r_{A_2},r'_{A_2}}$, where $r'_{A_1}$ and $r'_{A_2}$ are qudits used for purification.

Since only 1-bit classical communication is allowed, the general measurement performed at the node $C_1$ can be simulated by a two-outcome POVM described by $\{ \Pi_0, \Pi_1=\mathbb{I} - \Pi_0 \}$ on a two-qubitstate, which is a part of the entangled states denoted by $\ket{\Psi_{1}}_{ t_{A_1},r_{A_1},r'_{A_1}}$ and $\ket{\Psi_{2}}_ {t_{A_2},r_{A_2},r'_{A_2}}$.  
The implementability of $C_u$ over the collapsed butterfly network implies that the reduced state of qubits $r_{A_1}$, $r_{A_2}$ and qudits $r'_{A_1}$, $r'_{A_2}$ after the operation at node $C_1$ given by
\begin{equation}
\mathrm{Tr}_{t_{A_1},t_{A_2}} \left [\left ( \mathbb{I} \otimes \Pi_j \right) \ket{\Psi_{1}} \bra{\Psi_{1}} \otimes \ket{\Psi_{2}} \bra{\Psi_{2}} \right ]
\end{equation}
should be a pure state.  In order to obtain a two-qubitpure state by applying a general measurement on $\ket{\Psi_{1}} \bra{\Psi_{1}} \otimes \ket{\Psi_{2}} \bra{\Psi_{2}}$, the rank of each POVM element $\Pi_j$ ($j=0,1$) should be 1. But such a POVM does not exist for a two-outcome POVM on the four dimensional Hilbert space of two qubits. Therefore, our assumption was wrong and $F$ should be a 2-bit classical channel.

Finally, we show that the channels $G_1$ and $G_2$ should be 1-bit classical channels.  If we remove one of the channels $G_1$ and $G_2$, say $G_1$, the conditional operation at the node $B_1$ is no longer possible.  A measurement is performed at the node $C_2$ and the state after the measurement is changed depending on the outcome.  If the state after the measurement is not a maximally entangled state, it is not possible to transform the state of qubits at the nodes $B_1$ and $B_2$ to be a pure state by just performing operations at the node $B_1$.   Therefore, both of the channels $G_1$ and $G_2$ should be used as 1-bit classical channels and cannot be removed.

We note that if we constrain the inputs to be specific states, and the angle $\theta$ of $C_{u_\theta}$ to be a certain angle, use of one of the channels $G_1$ or $G_2$ is not necessary, while the other is used as a 2-bit classical channel. This happens when the state just before the final operations at the nodes $B_1$ and $B_2$ is maximally entangled and the operations at $B_1$ and $B_2$ are given by conditional unitary operations, since performing a local unitary operation $U_i$ on the one of the qubits of the maximally entangled state is equivalent to performing the transposed unitary operation $U_i^T$ on the other qubit.

\section{Summary and discussions} \label{Summary}

In this paper, in order to investigate distributed quantum computation under
restricted network resources, we introduce a quantum computation task over the
 butterfly network where both quantum and classical communications are limited.
  We have studied protocols implementing two-qubitunitary operations over a
 particular butterfly network introduced in \cite{Hayashi} by showing several
 constructions:
We have shown that unitary operations can be performed without adding any entanglement
 resource, if and only if the unitary operations are locally unitary equivalent to controlled unitary operations.  Our protocol is optimal in the sense that the unitary operations cannot be implemented if we relax the specifications of any of the channels.  We constructed a
 protocol for the case where one of the inner channels of the butterfly network
 is severely restricted in that it only allows one bit of classical information
 to be sent.   We also presented a modification of Hayashi's protocol that
 implements global Clifford operations over the butterfly network.

Our constructions show that by taking an appropriate coding, we can perform
 global unitary operations on spatially separated inputs and distribute the
outputs at the same time, even when restricted to a network where the quantum
channel connecting inputs and outputs is both directed and bottlenecked.
 Depending on the cost of resources in a given physical realization of the
 network, the coding varies.  In addition,  by studying the
 implementation of Clifford operations on the butterfly network, we also see
 the different characteristics of quantum and classical information, where the
 latter can be ``compressed'' and sent through the bottleneck whereas the former
 cannot.  In the bigger picture, results like these are the first steps toward developing a
 theory of \emph{network quantum resource inequalities}, which formalizes such
 tradeoffs in the more complicated network scenario, beyond the standard resource inequalities \cite{Min-Hsiu}.  

\begin{acknowledgments}
We thank T. Sugiyama and E. Wakakuwa for helpful discussions.  This work was supported by Special Coordination Funds for Promoting Science and Technology, MEXT, Japan and the Global COE Program, MEXT, Japan.
\end{acknowledgments}

\appendix*
\section*{Appendix: LEMMA 1 AND ITS PROOF}

The following lemma is required in the proof of necessity in Sec.~\ref{SecBoundCUnitary}.  The lemma and the proof are general for $d$-dimensional qudit systems, but in Sec.~\ref{SecBoundCUnitary}, only the case of $d=2$ is employed.

\begin{lemma}
If there exists a tensor product of a CPTP map on the Hilbert space $\mathcal{H}_{A_1} = \mathbb{C}^d \otimes \mathbb{C}^d$ at the node $A_1$ denoted by $\Lambda_{A_1}$ and another CPTP map on the Hilbert space $\mathcal{H}_{A_2} = \mathbb{C}^d \otimes \mathbb{C}^d$ at the node $A_2$ denoted by $\Lambda_{A_2}$  satisfying
\begin{eqnarray}
\left[ \Lambda_{A_1}  \otimes \Lambda_{A_2} \right]( \rho ) = \ket{\Theta} \bra{\Theta}
\label{ap-gammatransformation}
\end{eqnarray}
for $\rho \in \mathcal{S} (\mathcal{H}_{A_1} \otimes \mathcal{H}_{A_2})$ and $\ket{\Theta} \in \mathcal{H}_{A_1} \otimes \mathcal{H}_{A_2}$ given by
\begin{eqnarray}
\ket{\Theta}=\sum_{k=0}^{d^2-1} \sqrt{\lambda_k} \ket{\alpha_k} \ket{\beta_k},
\label{ap-finalschmidt}
\end{eqnarray}
where $\lambda_k\neq 0$ are Schmidt coefficients, and $\{ \ket{\alpha_k} \}$, $\{ \ket{\beta_k} \}$ are orthonormal bases of $\mathcal{H}_{A_1}$ and $\mathcal{H}_{A_2}$, respectively,  then $\rho$ must be a pure entangled state with Schmidt coefficients equal to those of $\ket{\Theta}$ with respect to the partition defined by $\mathcal{H}_{A_1} \otimes \mathcal{H}_{A_2}$.
\end{lemma}

{\it {Proof:}}  We denote the two qudits at the node $A_1$  by $t_1$, $r_1$, and the two qudits at the node $A_2$ by $t_2$ and $r_2$.    A decomposition of $\rho$ is given by $\rho=\sum_i p_i \ket{\varphi_i} \bra{\varphi_i}$, where $0 \leq p_i \leq 1$ for all $i$, $\sum_i p_i =1$, and $\{ \ket{\varphi_i} \}$ is a basis of $\mathcal{H}_{A_1} \otimes \mathcal{H}_{A_2}$.  Since
the right hand side  of Eq.~(\ref{ap-gammatransformation}) is a pure state, it must be that
\begin{eqnarray}
\left[ \Lambda_{A_1} \otimes \Lambda_{A_2} \right] (\ket{\varphi_i} \bra{\varphi_i}) = \ket{\Theta} \bra{\Theta}
\label{ap-gammaphitransformation}
\end{eqnarray}
for all $i$.  Since $\ket{\Theta}$ on the right hand side of Eq.~(\ref{ap-gammatransformation}) is a pure state with Schmidt number $d^2$, which is maximal for a bipartite cut of a 4-qudit entangled state, and the CPTP map $\Lambda_{A_1}  \otimes \Lambda_{A_2}$ cannot increase the Schmidt number, each state $\ket{\varphi_i} \in \mathcal{H}_{A_1} \otimes \mathcal{H}_{A_2}$ must be a pure state with Schmidt number $d^2$.

Using a Stinespring dilation, we rewrite the CPTP map $\Lambda_{A_1}  \otimes \Lambda_{A_2}$ by adding ancilla qudits $r'_1$ in a state $\ket{\xi_1} \in \mathcal{H}'_{A_1} = \mathbb{C}^{d'}$ at node $A_1$ and another ancilla qudit $r'_2$ in state $\ket{\xi_2} \in \mathcal{H}'_{A_2} = \mathbb{C}^{d'}$ at the node $A_2$, performing unitary operations $V_{A_1}$ on $\mathcal{H}_{A_1} \otimes \mathcal{H}'_{A_1}$
and  $V_{A_2}$ on $\mathcal{H}_{A_2} \otimes \mathcal{H}'_{A_2}$, and tracing out the ancilla degrees of freedom, (it is enough to consider  $d^2$-level ancillas).  Then Eq.~(\ref{ap-gammaphitransformation}) is transformed  to
\begin{eqnarray}
&\mathrm{Tr}_{r'_1, r'_2}  V_{A_1} \otimes V_{A_2}  \left( \ket{\varphi_i} \bra{\varphi_i}  \otimes \ket{\xi_1}  \bra{\xi_1} \otimes \ket{\xi_2}  \bra{\xi_2} \right )  V_{A_1}^\dag \otimes V_{A_2}^\dag \nonumber \\
&= \ket{\Theta} \bra{\Theta}.
\label{ap-gammaunitaryrepresentation1}
\end{eqnarray}
This means that before performing partial trace operations in Eq.~(\ref{ap-gammaunitaryrepresentation1}), the following relation should be satisfied
\begin{eqnarray}
V_{A_1} \otimes V_{A_2}  \ket{\varphi_i} \ket{\xi_1} \ket{\xi_2}
= \ket{\Theta}\ket{\Xi_i},
\label{ap-gammaunitaryrepresentation}
\end{eqnarray}
where $\ket{\Xi_i} \in \mathcal{H}'_{A_1} \otimes \mathcal{H}'_{A_2}$.

We have already shown that each $\ket{\varphi_i}$ is an entangled state with Schmidt number  $d^2$ (in terms of the systems $A_1$ and $A_2$). Since the state of the two ancilla qudits $\ket{\xi_1}\ket{\xi_2} $ in the left hand side  of Eq. (\ref{ap-gammaunitaryrepresentation})  is separable, and $V_{A_1} \otimes V_{A_2}$ is a separable unitary operation, the right hand side of Eq.~(\ref{ap-gammaunitaryrepresentation}) must have Schmidt number $d^2$, and thus, $\ket{\Xi_i}$ should be a pure product state.  We denote $\ket{\Xi_i} = \ket{c_i} \ket{d_i}$.   Because the Schmidt coefficients $\lambda_k$ are invariant under $V_{A_1} \otimes V_{A_2}$, the Schmidt coefficients of $\ket{\varphi_i}$  should be equal to those of $\ket{\Theta}$.   Thus we denote the Schmidt decomposition of $\ket{\varphi_i}$ using the same Schmidt coefficients defined in Eq.~(\ref{ap-finalschmidt}) by
\begin{eqnarray}
\ket{\varphi_i}=\sum_{k=0}^{d^2-1} \sqrt{\lambda_k}  \ket{a^i_k} \ket{b^i_k}
\label{ap-phischmidt}
\end{eqnarray}
where $\{ \ket{a^i_k} \}$ and $\{ \ket{b^i_k} \}$ are bases of $\mathcal{H}_{A_1}$ and $\mathcal{H}_{A_2}$, respectively, for all $i$.

We consider the state at node $A_1$ obtained by the partial trace of the state given by Eq.~(\ref{ap-gammaunitaryrepresentation}) over the systems $t_2$, $r_2$ and $r'_2$ at node $A_2$.   By using the Schmidt decompositions given by Eqs.~(\ref{ap-finalschmidt}) and (\ref{ap-phischmidt}), we obtain
\begin{eqnarray}
V_{A_1} \left ( \sum_k \lambda_k \ket{a^i_k}\bra{a^i_k} \otimes \ket{\xi_1} \bra{\xi_1} \right ) V^\dagger_{A_1} \nonumber \\
= \sum_k \lambda_k \ket{\alpha_k} \bra{\alpha_k} \otimes \ket{c_i} \bra{c_i}.
\label{ap-tracingouta2}
\end{eqnarray}

To handle any degeneracy of the Schmidt coefficients, we rewrite the bases $\{ \ket{a^i_k} \}$ and $\{ \ket{\alpha_k} \}$ by $\{ \ket{a^i_{l,{m}}} \}$ and $\{ \ket{\alpha_{l,m}} \}$, respectively, where $l$ specifies the value of the Schmidt coefficient, and $m$ specifies the degeneracy.   Using these notations, the identity operators for the Hilbert space
$\mathcal{H}_{A_1}$ can be written by
\begin{eqnarray}
\mathbb{I}_{\mathcal{H}_{A_1}}=\sum_k \ket{a^i_k} \bra{a^i_k} = \sum_l \sum_m \ket{a^i_{l,m}} \bra{a^i_{l,m}} ,
\label{ap-identity}
\end{eqnarray}
for all $i$, and also
\begin{eqnarray}
\mathbb{I}_{\mathcal{H}_{A_1}}=\sum_k \ket{\alpha_k} \bra{\alpha_k} = \sum_l \sum_m \ket{\alpha_{l,m}} \bra{\alpha_{l,m}}.
\end{eqnarray}

In this new notation, Eq.~(\ref{ap-tracingouta2}) is written
\begin{eqnarray}
V_{A_1} \left ( \sum_l  \lambda_l  \sum_m \ket{a^i_{l,m}}\bra{a^i_{l,m}} \otimes \ket{\xi_1} \bra{\xi_1} \right ) V^\dagger_{A_1} \nonumber \\
= \sum_l \lambda_l \sum_m  \ket{\alpha_{l,m}} \bra{\alpha_{l,m}} \otimes \ket{c_i} \bra{c_i}.
\end{eqnarray}
Since $\{ \ket{a^i_{l,m}} \}$ and $\{ \ket{\alpha_{l,m}} \}$ are bases and $\lambda_l \neq 0$ is guaranteed for all $l$,  we obtain
\begin{eqnarray}
V_{A_1} \left ( \sum_m \ket{a^i_{l,m}}\bra{a^i_{l,m}} \otimes \ket{\xi_1} \bra{\xi_1} \right ) V^\dagger_{A_1} \nonumber \\
= \sum_m  \ket{\alpha_{l,m}} \bra{\alpha_{l,m}} \otimes \ket{c_i} \bra{c_i}.
\label{ap-usefulrelation}
\end{eqnarray}

Now we consider an operator $K$ independent of the index $i$ defined by
\begin{eqnarray}
K= V_{A_1} \left ( \mathbb{I}_{\mathcal{H}_{A_1}} \otimes \ket{\xi_1} \bra{\xi_1} \right ) V^\dagger_{A_1} .
\end{eqnarray}
Using Eqs. (\ref{ap-identity}) and (\ref{ap-usefulrelation}), we have
\begin{eqnarray}
&K= V_{A_1} \left ( \sum_l \sum_m \ket{a^i_{l,m}} \bra{a^i_{l,m}} \otimes \ket{\xi_1} \bra{\xi_1} \right ) V^\dagger_{A_1} \nonumber \\
&= \sum_l \sum_m  \ket{\alpha_{l,m}} \bra{\alpha_{l,m}} \otimes \ket{c_i} \bra{c_i} \nonumber \\
&= \mathbb{I}_{\mathcal{H}_{A_1}} \otimes \ket{c_i} \bra{c_i} .
\label{ap-usefulformula}
\end{eqnarray}
The operator $K$ should be independent of the index $i$, and therefore, we can take $\ket{c_i}=e^{i \delta'_i} \ket{c}$ where $\delta'_i$ denotes an $i$-dependent phase factor.   Arguing similarly for the state at node $A_2$ obtained as a partial trace of the state given by
by Eq.~(\ref{ap-gammaunitaryrepresentation}), we can conclude that $\ket{d_i}=e^{i \delta''_i} \ket{d}$.  Thus, Eq.~(\ref{ap-gammaunitaryrepresentation}) is now given by
\begin{eqnarray}
V_{A_1} \otimes V_{A_2}  \ket{\varphi_i} \ket{\xi_1} \ket{\xi_2}
= e^{i \delta_i} \ket{\Theta}\ket{\Xi},
\label{ap-gammaunitaryrepresentation2}
\end{eqnarray}
where $\delta_i = \delta'_i + \delta''_i$.

Taking the inner product of $V_{A_1} \otimes V_{A_2}  \ket{\varphi_i} \ket{\xi_1} \ket{\xi_2}$ and $V_{A_1} \otimes V_{A_2}  \ket{\varphi_j} \ket{\xi_1} \ket{\xi_2}$ for $j \neq i$, we obtain $\bra{\varphi_j} \varphi_i \rangle =e^{i \left ( \delta_{i} - \delta_{j} \right )}$.  This relation indicates that $\ket{\varphi_i} \bra{\varphi_i} =\ket{\varphi_j} \bra{\varphi_j} $ for all $i$ and $j$.  Hence, $\rho=\sum_i p_i \ket{\varphi_i} \bra{\varphi_i}$ turns out to be a mixture of the same state, and $\rho=\ket{\varphi}\bra{\varphi}$ is a pure state with Schmidt coefficients equal to those of $\ket{\Theta}$.


\begin{thebibliography}{99}
\bibitem{Hayashi}
	M. Hayashi, Phys. Rev. A \textbf{76}, 040301(R) (2007).
\bibitem{DQC}
	H. Buhrman and H. R\"{o}hrig, Lecture Notes in Computer Science, \textbf{2747}, 1 (2003).
\bibitem{QComC} I. Kremer, Master's thesis, 
	Hebrew University of Jerusalem, 1995.
\bibitem{Qi:200}
	T. M. Cover and J. A. Thomas, {\it Elements of
	Information Theory} (John Wiley \& Sons, Hoboken, 2006).
\bibitem{Ajlswede} 
	R. Ahlswede {\it et. al.}, IEEE Trans. Inf. Theory \textbf{46}, 1204 (2000).
\bibitem{Hayashi et al}
	M. Hayashi {\it et. al.}, Lecture Notes in Computer Science \textbf{4393}, 610 (2007).
\bibitem{Debbie}
	D. Leung and J. Oppenheim and A. Winter, IEEE Trans. Inf. Theory, \textbf{56} (2010) 3478.
\bibitem{Kobayashi}
	H. Kobayashi {\it et. al.}, Lecture Notes in Computer Science \textbf{5555}, 622 (2009).
\bibitem{Min-Hsiu}
	C. H. Bennett, Quantum Information and Computation \textbf{4}, 460 (2004);
	I. Devetak, A.W. Harrow and A. Winter, Phys. Rev. Lett. \textbf{93}, 230504 (2004);
	IEEE Trans. Inf. Theory \textbf{54}, 4587 (2008);
	M.-H. Hsieh and M. M. Wilde, IEEE Trans. Inf. Theory \textbf{56}, 4705 (2010).
\bibitem{Superdensecoding}C. H. Bennett and S. J. Wiesner, Phys. Rev. Lett.
	\textbf{69}, 2881 (1992).
\bibitem{teleportation}
	C. H. Bennett {\it et. al.}, Phys. Rev. Lett. \textbf{70},
 1895 (1993).
\bibitem{entanglingpower} B. Kraus and J. I. Cirac, Phys. Rev. A \textbf{63},  062309 (2001).
\bibitem{delocalizationpower} A. Soeda and M. Murao, New J. Phys. \textbf{12}, 093013 (2010).
\bibitem{LOCC} J. Eisert, K. Jacobs, P. Papadopoulos and M. B. Plenio, Phys. Rev. A \textbf{62}, 052317 (2000).
\bibitem{SchmidtDecomp} W. D\"{u}r, G. Vidal and J.I. Cirac, Phys. Rev. Lett. {\bf 89} 057901 (2002).
\bibitem{comment}  We choose the phrase ``entangling capability" to avoid confusion with ``entangling power" \cite{entanglingpower}, which is conventionally used to evaluate entangling capability particularly in terms of the maximal amount of entanglement generated by operations.
\bibitem{instrument} E. B. Davies, J. T. Lewis, Comm. Math. Phys. {\bf 17}, 239 (1970); M. Ozawa, J. Math. Phys. {\bf 25}, 79 (1984).
\bibitem{SLOCC} G. Vidal, Phys. Rev. Lett. {\bf 83}, 1046 (1999).
\bibitem{LOCCimplementation} A. Soeda, P.S. Turner and M. Murao, arXiv: 1008.1128 (2010).
\end{thebibliography}
\end{document}